\documentclass[aps,prl,twocolumn,nofootinbib,preprintnumbers,superscriptaddress]{revtex4}
\usepackage{amssymb,amsmath,amsbsy}
\usepackage{graphicx,color}
\definecolor{purple}{rgb}{0.7,0.0,0.5}
\pdfoutput=1

\newcommand{\be}{\begin{equation}}
\newcommand{\ee}{\end{equation}}
\newcommand{\bea}{\begin{eqnarray}}
\newcommand{\eea}{\end{eqnarray}}
\newcommand{\lb}{\left(}
\newcommand{\rb}{\right)}
\newcommand{\la}{\langle}
\newcommand{\ep}{\epsilon}
\newcommand{\ra}{\rangle}
\newcommand{\nn}{\nonumber}
\newcommand{\p}{\partial}
\newcommand{\mO}{{\mathcal O}}

\newcommand{\nbox}{{\,\lower0.9pt\vbox{\hrule \hbox{\vrule height 0.2 cm \hskip 0.19 cm \vrule height 0.2 cm}\hrule}\,}}
\newcommand{\Tr}{\ {\rm Tr}}

\def\href#1#2{#2}

\newcommand{\bi}{\begin{itemize}}
\newcommand{\ei}{\end{itemize}}
\newcommand{\ben}{\begin{enumerate}}
\newcommand{\een}{\end{enumerate}}
\newcommand{\bca}{\begin{cases}}
\newcommand{\eca}{\end{cases}}

\newcommand\vev[1]{{\ensuremath{\left\langle{#1}\right\rangle}}}

\newcommand\ket[1]{\ensuremath{\lvert{#1}\rangle}}
\newcommand\bra[1]{\ensuremath{\langle{#1}\rvert}}

\newcommand{\cR}{\mathcal{R}}

\begin{document}

\title{Eigenstate Thermalization Hypothesis in Conformal Field Theory}

\author{Nima Lashkari}
\affiliation{Center for Theoretical Physics, Massachusetts Institute of Technology, Cambridge, MA 02139}

\author{Anatoly Dymarsky}
\affiliation{Department of Physics and Astronomy, University of Kentucky, Lexington, KY 40506, USA\\ Skolkovo Institute of Science and Technology, Skolkovo Innovation Center, Moscow 143026 Russia}

\author{Hong Liu}
\affiliation{Center for Theoretical Physics, Massachusetts Institute of Technology, Cambridge, MA 02139}

\begin{abstract}
We investigate the eigenstate thermalization hypothesis (ETH) in d+1 dimensional conformal field theories by studying reduced density matrices in energy eigenstates. We show that if local probes of high energy primary eigenstates satisfy ETH, then any finite energy observable with support on a subsystem of finite size satisfies ETH. In two dimensions, we discover that if ETH holds locally, the 
finite size reduced density matrix of states created by heavy primary operators is well-approximated by a projection to the Virasoro identity block. 

%
\end{abstract}
\maketitle

\vskip 1cm

\section{Introduction}

Whether and how an isolated interacting quantum system initially out of equilibrium equilibrates is a complex dynamical 
question about which we currently have little analytic control.  
For a non-integrable system, there is nevertheless a powerful conjecture--the Eigenstate Thermalization Hypothesis (ETH)--which states that expectation values of generic few-body observables in a finitely excited energy eigenstate should coincide with those in the micro-canonical ensemble up to corrections that are exponentially small in entropy~\cite{Srednicki:1995pt,Deutsch:1991}.  Assuming ETH, one can then readily deduce that in the thermodynamic limit: (i) expectation values of few-body observables in a generic state coincide with those in a thermal ensemble; (ii)  expectation values of few-body observables in a non-equilibrium initial state will evolve towards those of a thermal ensemble. 

By now, there are many numerical evidences for ETH in a variety of quantum systems~\cite{Rigol:2008}. While ETH was often stated for few-body observables, recently numerical supports were also found for the full reduced density matrix of any subsystem~\cite{Garrison:2015lva,Dymarsky:2015cxp}.

ETH is powerful as instead of following time evolution of a general non-equilibrium state, we now face an in-principle  simpler problem of understanding properties of finitely excited energy eigenstates. 
Unfortunately, despite much numerical progress energy eigenstates of a general interacting many-body system are still too complex to be tractable by analytic methods. A proof of ETH for general non-integrable systems appears out of reach at this stage.  

In this paper we show that progress can be made in conformal field theories (CFT). 
We are able to prove that for CFTs, a weakest form ETH which applies to local {\it primary} operators (to which we refer as local ETH) in fact leads to a strongest form of ETH which applies to the reduced density matrix of a general subsystem. 
Thus to prove ETH for CFTs it is enough to prove that it applies to local primary operators which in turn reduces 
to a statement regarding coefficients in operator product expansions (OPE). Clearly not all CFTs satisfy ETH, thus proving local ETH should amount to understanding how the corresponding statement regarding OPE coefficients connects with non-integrability. 

We believe that local ETH provides a powerful 
technical handle for obtaining many dynamical properties of a CFT. As a simple application 
we use it to compute the Renyi entropies for an interval in an energy eigenstate in a (1+1)-dimensional CFT. Interestingly we find that even though the reduced density matrix for the region approaches that of the canonical ensemble in the thermodynamic
limit, the Renyi entropies differ by a finite amount.

Our discussion should be generalizable to general quantum field theories which we will pursue elsewhere.

\section{Setup}



Consider finitely excited energy eigenstate $\ket{E}$ of a conformal field theory in a $d$-dimensional space of finite volume. For simplicity, we take it to be a sphere of radius $L$ (i.e. the total spacetime is $\mathbb{R}_t \times S^d$).
A key simplification of conformal symmetry is that in Euclidean signature any state $\ket{\psi}$ on $S^d$ can be mapped through 
conformal transformation 
\bea \label{c1}
&&ds^2_{\rm cylinder}=d\tau^2+L^2d\Omega_d^2 
={\Lambda^2} (dr^2+r^2d\Omega_d^2)\nn\\
&&\Lambda=\frac{L}{r},\qquad \tau=L\log r 
\eea
to a {\it local} operator $\Psi$ inserted at origin ($r=0$) in $\mathbb{R}^{d+1}$. Similarly, the conjugate state 
$\la \psi|$ is mapped to a local operator $\Psi^\dagger$ inserted at $r=\infty$ in $\mathbb{R}^{d+1}$. Energy eigenstates are in one-to-one correspondence to the operators $P_{\mu_1}\cdots P_{\mu_m}\Psi_a$, where $P_{\mu}$ is a momentum operator and $\Psi_a$ is a primary operator. Below, we will use  state and operator languages interchangeably. 

Kinematics of conformal symmetry fixes the correlators in descendant states $P_{\mu_1}\cdots P_{\mu_m}\Psi_a$, in terms of those of their corresponding primary $\Psi_a$. The dynamical content of a CFT is the spectrum of its primaries and OPE coefficients. As a dynamical statement about energy eigenstates which are not related by symmetries 
ETH 
should be restricted to primary energy eigenstates. 
Moreover, we will restrict our attention to homogeneous energy eigenstates on $S^d$ which are invariant under $SO(d+1)$ rotations. Such an energy eigenstate with energy $E_a$ corresponds to a spinless primary $\Psi_a$ of dimension $h_a = E_a/L$. The energy density of the system is $\ep_a = {E_a \over L^d \omega_d} = {h_a \over L^{d+1} \\
_d}$, where $\omega_d$ is the volume of a unit $S^d$. For a CFT in a 
 thermal state of temperature $T$, $\ep_a \sim C T^{d+1}$ where $C$ is 
central charge, which motivates us to define a ``thermal'' length scale associated with $\ket{E_a}$
 \be 
 \lambda_T=\left({\ep_a \over C}\right)^{-{1 \over d+1}} \sim T^{-1} \ .
\ee
We will be interested in the thermodynamic limit with $L \to \infty$, while keeping a finite energy density $\ep_a$ and thus a finite 
$\lambda_T$. 
In this limit the scaling dimension $h_a$ should then scale with $L$ as 
\be\label{1ex}
h_a = C \omega_d \left({L \over \lambda_T}\right)^{d+1} \ .
\ee


 The {\it local ETH} condition is defined as 
\be\label{ETH}
\la E_a|\mO_p|E_b\ra= O_p (E)\delta_{ab}+ O_{ab}^p,
\ee
where $\mO_p$ is a local primary operator (with $p$ labeling different operators), the diagonal element $O_p (E)$ is a smooth function of $E = {E_a + E_b \over 2}$, and  $O_{ab}^p \sim  e^{- O(S(E))}$ with $e^{S (E)}$ the density of state at energy $E$. 
Due to homogeneity, it does not matter where $\mO_p$ is inserted, so we left the location of $\mO_p$ implicit. 
In equation~\eqref{ETH} $\mO_p$ should be understood as a smeared operator $\mO_p (f)=\int_{S^d}f \mO_p$ with smearing function $f$ such that the set of observables $\{\mO_p (f)\}$ form an algebra. In particular, the operator $\mO^2_p$ should be well-defined and has finite expectation values in physical states\footnote{Note that for an expectation value of $\mO$ to be physically meaningful, its fluctuations should be sufficiently small. This implies that the corresponding expectation value of $\mO^2$ must be finite. Thus smeared operators must be used in formulating ETH in quantum field theory.}. Below for notational simplicity we will continue to suppress $f$ in $\mO_p (f)$.

We now introduce {\it subsystem ETH} which states that for a subregion $B$~\cite{Dymarsky:2015cxp} 
\ben 
\item There exists a ``universal'' density matrix $\rho_B (E)$ (which depends only  on $B$ and energy $E$) 
such that for any energy eigenstate $\ket{E_a}$
\be \label{cmeth}
||\rho_B^a - \rho_B (E = E_a) ||  \sim e^{-O(S(E))}
\ee
where $\rho_B^a \equiv \Tr_{B^c} \ket{E_a} \bra{E_a}$. In~\eqref{cmeth}, $\|\rho-\sigma\| =\frac{1}{2}{\rm Tr}\sqrt{(\rho-\sigma)^2}=\frac{1}{2}\sum_i|\lambda_i|
$ is  the trace distance 
of two density matrices $\rho$ and $\sigma$, and $\lambda_i$ are the eigenvalues of the Hermitian operator $\rho-\sigma$.

\item Introducing $\sigma_{ab} \equiv  \Tr_{B^c} \ket{E_a} \bra{E_b} $, then 
 \be \label{dmeth}
\|e^{i\alpha}\sigma_{ab}+e^{-i\alpha}\sigma_{ba} \| \sim e^{-O\lb S(E)\rb}  \ .
\ee
for all $\alpha$. 
\een

Note that subsystem ETH~\eqref{cmeth}--\eqref{dmeth} implies local ETH, that is, for an operator $\mO$ supported 
inside region $B$ we have
\be \label{j1}
\Tr((\rho_B^a-\rho_B(E)\mO)\leq 
\|\rho_B^a-\rho_B(E)\|^{1 \over 2} \Tr((\rho_B^a+\rho_B(E))\mO^2)^{1\over 2}
\ee
and thus is exponentially small. Similarly, $\Tr (\mO \sigma_{ab} )=e^{-O(S(E))}$.
See supplementary material for a proof of~\eqref{j1}.\footnote{Expectation values depend on the normalization of the observable and could be arbitrarily large for unbounded operators. It is important to appreciate that density matrices with small trace distance are physically indistinguishable, even though corresponding expectation values of certain operators might not be close \cite{Lashkari:2014pna}.}

To close our setup, let us briefly comment on the descendant eigenstates. As an example consider $\ket{E}$ corresponding to a spinless operator $(P^2_\mu)^l \Psi_a$ where  $l$ an integer. For such a state $E = E_{a,l}= E_a + {2l \over L}$.  The matrix element $\vev{E|\mO_p|E}$ is controlled by that in the primary state $\Psi_a$, i.e. by $O_p (E - {2l \over L})$ rather than $O_p (E)$.  
Thus as stated earlier, we should not include the descendant states in either~\eqref{ETH} or~\eqref{cmeth}--\eqref{dmeth}.
Of course if one assumes~\eqref{ETH}, for $h_a \gg l$, from smoothness of function $O_p (E)$, $O_p (E - {2l \over L})$ is related to $O_p (E)$ only by corrections of order $l/h_a$ and similarly the corresponding reduced density matrix is close to  
$\rho_B (E)$ in trace distance with  corrections that are polynomially suppressed in $l/h_a$. But such statements do not reflect any new dynamics beyond~\eqref{ETH}--\eqref{dmeth}.
For completeness, in supplementary material we discuss more explicitly the story for descendant states.

\section{From local ETH to subsystem ETH}

Let us now look at the implications of~\eqref{ETH}. 
Using conformal mapping~\eqref{c1} the matrix element $\la E_b|\mO_p|E_a\ra$ on $S^d$ is mapped to the Euclidean three-point function
\bea\label{onePoint}
&&\la E_b|\mO_p|E_a\ra \\
&&=\frac{L^{-h_p}\la\Psi_{b}^\dagger (\infty)\mO_p(1)\Psi_{a}(0)\ra}{\sqrt{\la \Psi_{a}^\dagger (\infty)\Psi_{a}(0)\ra\la \Psi_{b}^\dagger (\infty)\Psi_{b}(0)\ra}} \nn
=C^p_{ab}L^{-h_p},
\eea
where $C^p_{ab}$ is the OPE coefficient for $\mO_p$ appearing in the operator product $\Psi_a$ and $ \Psi_b^\dagger$, when $\Psi_a$ and $\Psi_b$ are primaries. 
We assume the thermodynamic limit exists in~\eqref{ETH} for any operator $\mO_p$ whose dimension $h_p$ does not scale with $L$.
From~\eqref{onePoint} and expressing $L$ in term of $h_a$ using~\eqref{1ex} we conclude that 
the OPE coefficient $C_{ab}^p$ must scale with $h_a \to \infty$ as
\be \label{eo}
C_{ab}^p 
= h_a^{h_p\over d+1} \delta_{ab} f_p (E) + \cR_{ab}^p 
\ee
where $f_p (E) =  \lambda_T^{h_p} (N \omega_d)^{- {h_p\over d+1}} O_p (E) $ is a smooth function of $E$ independent of label $a$, and $\cR_{ab}^p = L^{h_p} O_{ab}^p \sim e^{- O(h_a^{d \over d+1}) + {h_p \over d+1} \log h_a} $ as on general ground we expect that 
$S(E) \propto L^d$ in the thermodynamic limit. Note that equation~\eqref{eo} implies the following: for operators $\mO_p$
whose $C_{aa}^p$ grow slower than $h_a^{h_p\over d+1}$ with $h_a$ cannot have a non-vanishing expectation value in the thermodynamic limit, while it is impossible for an operator with $C_{ab}^p$ to grow faster than $h_a^{h_p\over d+1}$ as that would imply the thermodynamic limit does not exist. 


\begin{figure}
\centering
\includegraphics[width=0.4\textwidth]{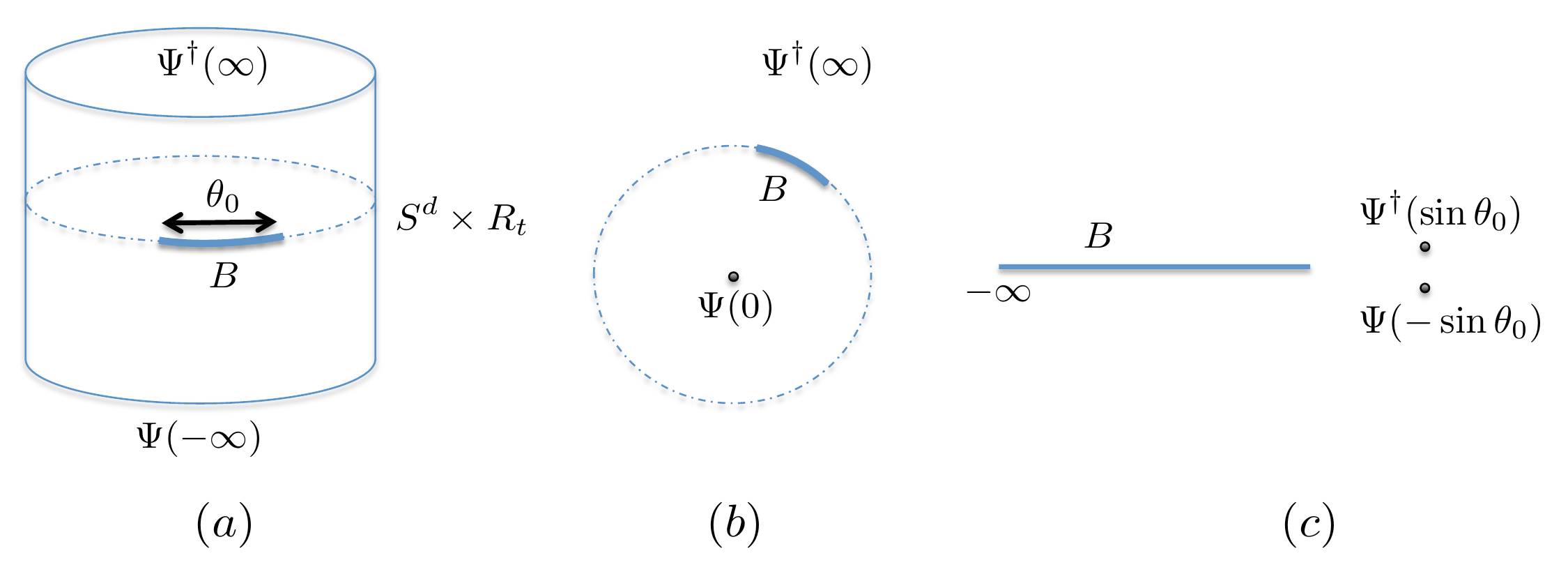}\\
\caption{\small{(a) The cylinder $S^d\times R_t$ conformal frame (b) Radial quantization $\mathbb{R}^{d+1}$ conformal frame. (c) The conformal frame convenient for the study of the density matrix on subsystem $B$.}}
\label{fig2}
\end{figure}

We are interested in reduced density matrices of a region $B$ of a finite size in the thermodynamic limit.
Consider $B$ to be a ball around the north pole of $S^d$ with angular radius $\theta_0=\tan^{-1}(\frac{R}{2L})$. 
In the the thermodynamic limit of the angular size $R/L\to 0$ keeping $R$ finite, $B$ becomes a ball of radius $R$ in $\mathbb{R}^d$. We will discuss other shapes or regions with disconnected components toward the end. 

The trace distance is invariant under unitary rotation of states:
\bea
\|\rho-\sigma\|=\|\tilde{\rho}-\tilde{\sigma}\|,\qquad \tilde{A}=U A U^\dagger.\nn
\eea
Therefore, we are free to compare density matrices in any conformal frame. A convenient conformal frame is the Rindler frame
\be
ds^2_{\rm cylinder}=\Lambda^2(X) dX^idX^i,
\ee
where the ball-shaped region $B$ on $S^d$ is mapped to negative half-space, $X^1\leq 0$; see figure \ref{fig2} and 
supplementary material for details. Under this map $\Psi_a$ and $\Psi_a^\dagger$ that create the state vector and its dual are mapped to $X^\mu_\pm = (\pm \sin\theta_0, \cos\theta_0, X^{i>1}=0)$, respectively. Thus, in the thermodynamic limit $\theta_0 \to 0$, the two operators are colliding in this frame, and we can use their operator product expansions to represent them. 
%

More explicitly, denote the reduced density matrix for region $B$ coming from the energy eigenstate $|E_a\ra$ in the Rindler frame with $\tilde{\rho}_B^a$. The expectation values in this state can be written as 
\bea
{\rm Tr} (\tilde{\rho}_B^a \cdots) &= & \frac{\la \Psi^\dagger_a (X^\mu_+)\Psi_a (X^\mu_- )\cdots\ra}{\la \Psi^\dagger_a (X^\mu_+)\Psi_a (X^\mu_- )\ra} \cr
&=& \sum_{h_p} C^{p}_{aa } \left({ R \over L} \right)^{h_p}\la \mO_{p} (X_0) \cdots \ra
\label{ene}
\eea
where $X_0^\mu = (0,1, X^{i > 1} =0)$. Note that the reduced density matrix can be considered  a map from the algebra of observables on the subsystem to expectation values: $\rho(\mO\in \mathcal \{\mO\})={\rm Tr} (\rho \mO)$. For subregion $B$, we should restrict to operators $\mO_p$ whose dimensions $h_p$ remain finite in the thermodynamic limit (i.e. does not scale $L$). The summation in~\eqref{ene} should be understood as so. Also, it should be understood that in the sum in~\eqref{ene} the descendants of $\mO_p$ are included implicitly (similarly below). 

\begin{figure}
\centering
\includegraphics[width=0.4\textwidth]{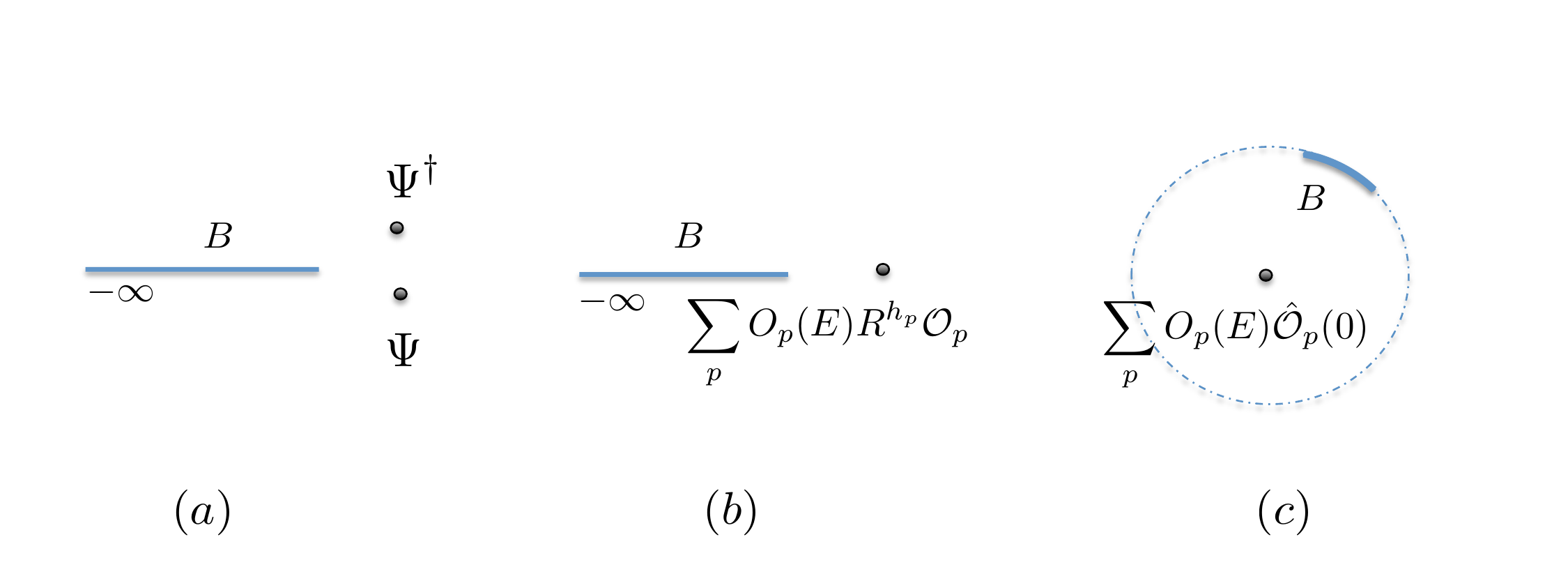}\\
\caption{\small{ (a) and (b) The Rindler frame path-integrals that prepare, respectively, $\tilde{\rho}_B^a$ and $\tilde{\rho}_B(E)$ (c) The path-integral for $\rho_B(E)$ in the radial quantization frame.}}
\label{fig2p}
\end{figure}

Using~\eqref{onePoint} and~\eqref{ETH} we can write~\eqref{ene} in an operator form 
\be\label{11}
\tilde{\rho}_B^a =  \tilde{\rho}_B (E)+  \cR_B^a 
\ee
where 
\bea \label{rhouniv}
\tilde{\rho}_B (E) &=&  \sum_{h_p} O_p (E)  R^{h_p} \mO_{p} (X_0) \\
\cR_B^a  &=&  \sum_{h_p} O_{aa}^p  R^{h_p} \mO_{p} (X_0)  \ .
\eea
Equation~\eqref{rhouniv} (and similarly for $\cR_B^a$) should be understood as follows. 
 $\tilde{\rho}_B (E)$ is prepared in the Rindler frame via a Euclidean path-integrals over $\mathbb{R}^{d+1}$ with boundary conditions above and below the negative half-space and a local operator inserted at $X_0^\mu$ (see figure \ref{fig2p}). In~\eqref{rhouniv} we denote the density matrix by the specific operator inserted at $X_0^\mu$.
 Similarly we can write $\sigma_{ab}$ introduced  
before~\eqref{dmeth} as (with $a \neq b$)
\be \label{12}
\tilde{\sigma}_{ab} =  \sum_{h_p} O_{ab}^p  R^{h_p} \mO_{p} (X_0)  \ .
\ee
The equations above are to be understood as operator equalities inside Euclidean path-integral in the Rindler frame. Mapping back to the radial quantization frame the universal density matrix $\rho_B(E)$ is
\bea
&&\rho_B(E)=\sum_{h_p}O_p(E) R^{h_p} \hat{O}_p(0)\nn\\
&&\hat{O}_p=U^\dagger \mO_p U,
\eea
and $U$ is the unitarity corresponding to the conformal transformation from the Rindler frame to radial quantization frame; see figure \ref{fig2p}(c).

From~\eqref{11} and the finite radius of convergence of OPE we immediately find that for observables $\mO_1, \cdots \mO_k$ in region $B$, 
\bea
&&\frac{\Delta \la \mO_1\cdots \mO_k\ra}{\la \mO_1\cdots \mO_k\ra}\equiv \frac{{\rm Tr} ((\tilde{\rho}_B^a -\tilde{\rho}_B)\mO_1\cdots \mO_k)}{{\rm Tr}  (\tilde\rho_B\mO_1\cdots \mO_k)}\nn\\
&&=\frac{\sum_{h_p} O_{aa}^p R^{h_p}\la \mO_{p}\mO_1\cdots \mO_k\ra}{\sum_{h_p} \mO_p (E) R^{h_p}\la \mO_{p}\mO_1\cdots \mO_k\ra} \sim e^{- O(S(E))}\ .\nn
\eea
In fact, we can prove that density matrices $\rho_B^a$ and $\rho_B(E)$ are close in trace distance. Trace distance is hard to compute in continuum theories. Instead, we use another measure of distance called relative entropy which according to Pinsker's inequality provides an upper bound on trace distance of nearby states: 
\bea\label{pinsker}
\|\rho_B^a-\rho_B(E)\|^2=\|\tilde{\rho}_B^a-\tilde{\rho}_B(E)\|^2\leq 2S(\tilde{\rho}^a_B\|\tilde{\rho}_B(E)) \ .\nn
\eea
Since the two density matrices are close, we only need to compute relative entropy $S(\tilde{\rho}^a_B\|\tilde{\rho}_B)$ perturbatively in small $\cR^a_B$. To the second order, we find that the relative entropy is given by the quantum Fisher information
\bea\label{mainresult}
&&S(\tilde{\rho}^a_B\|\tilde{\rho}_B)\simeq\int\limits_0^\infty \frac{ds}{2}\Tr\lb (\tilde{\rho}_B+s)^{-1}\cR^a_B(\tilde{\rho}_B+s)^{-1}\cR^a_B\rb\nn\\
&&=O(\eta_a^2),\nn\\
&&\eta_a=\sup_p O^p_{aa}=e^{-O(S(E))},
\eea
where we have used the integral representation of the logarithm of a positive operator. In supplementary material, we expand relative entropy to all orders in $\eta_a$ using the replica trick in  \cite{Lashkari:2015dia}, and show (\ref{mainresult}) assuming that the $n$-point correlators on an $n$-sheeted manifold are finite.
%
%
Invoking (\ref{pinsker}) it is evident that
\bea
\|\rho_B^a-\rho_B(E)\|^2\leq 2S(\tilde{\rho}_B\|\tilde{\rho}_B(E))=O(\eta_a^2),
\eea
which demonstrates the subsystem ETH. Note that subsystem ETH holds for any finite ball-shaped subsystem of finite radius $R$. Monotonicity of trace distance under partial trace then implies that subsystem ETH holds for any subsystem of arbitrary shape and disconnected regions that can be encircled in a ball of finite size:
\bea
&&\forall A\qquad \text{such that}\qquad \exists B: \:A\subset B,\nn\\
&&\|\rho_A^a-\rho(E)_A\|\leq \|\rho_B^a-\rho(E)_B\|=O(\eta_a).
\eea
where $\rho_A(E)=\Tr_{B-A}\rho_B(E)$; see figure \ref{fig2p}(c). 

Now consider the state $|E_{a,b,\alpha}\ra=\frac{1}{\sqrt{2}}(|E_a\ra+e^{i\alpha}|E_b\ra$) with $|E_a\ra$ and $|E_b\ra$ two energy eigenstates. The reduced density matrix in this state in the Rindler frame is 
\bea
\tilde{\rho}^{a,b,\alpha}_B&=&\frac{1}{2}(\tilde{\rho}^a_B+\tilde{\rho}^b_B+e^{i\alpha}\tilde{\sigma}_{ab}+e^{-i\alpha}\tilde{\sigma}_{ba})\nn\\
&=&\frac{1}{2}\lb\tilde{\rho}_B(E_a)+\tilde{\rho}_B(E_b)\rb+\cR^{a,b,\alpha}_B,
\eea
where
\bea
\cR^{a,b,\alpha}=\frac{1}{2}\sum_{h_p}\lb O^p_{aa}+O^p_{bb}+e^{i\alpha}O^p_{ab}+e^{-i\alpha}O^p_{ba}\rb R^{h_p}\mO_p(X_0)\nn,
\eea
and we have used the local ETH assumption in (\ref{ETH}). Repeating the argument above one finds
\bea
&&\|e^{i\alpha}\sigma_{ab}+e^{-i\alpha}\sigma_{ba}\|=\|\tilde{\rho}_B^{a,b,\alpha}-\frac{1}{2}(\tilde{\rho}_B(E_a)+\tilde{\rho}_B(E_a))\|\nn\\
&&\leq S\lb\tilde{\rho}_B^{a,b,\alpha}\|\frac{1}{2}(\tilde{\rho}_B(E_a)+\tilde{\rho}_B(E_a))\rb=O(\eta_{ab}^2)\nn\\
&&\eta_{ab}=\sup_p\lb O^p_{aa}+O^p_{bb}+e^{i\alpha}O^p_{ab}+e^{-i\alpha}O^p_{ba}\rb=e^{-O(S(E))}.\nn
\eea
By a similar Euclidean path-integral argument, the following two-norms are also small
\bea
\sqrt{{\rm Tr}\lb (\rho_B^a-\rho_B(E))^2\rb}=O(\eta_a),\nn\\
\sqrt{{\rm Tr}\lb (\sigma_{ab}\sigma_{ab}^\dagger)^2\rb}=O(\eta_a).
\eea
However, an explicit computation of the two-norm requires the propagator on a two-sheeted manifold that we do not have know. 
     


\section{Further applications of local ETH}

The expression in (\ref{11}) is an operator equality. Expanding the Renyi entropies of the density matrix in the energy eigenstate it is easy to see that $$S_n(\rho^a_B)-S_n(\rho_B(E))=e^{-O(S(E))}.$$ That is to say that the Renyi entropies of reduced density matrices in energy eigenstates are universal. 

CFTs in $1+1$-dimensions are special in that the expectation value of local operators vanishes in the thermal state. 
That is due to the fact that the thermal state on a line is conformally flat. Then, the local ETH implies that in the thermodynamic limit $C^p_{aa}\rightarrow 0$ for all quasi-primaries, except for  those made of stress tensor $T$ and $\bar{T}$. In other words, the density matrices in eigenstates are well-approximated by their projection to the Virasoro identity block. Assuming local ETH one can compute the universal density matrix $\tilde{\rho}_B(E)$ defined by (\ref{rhouniv}). Since the quasi-primaries that appear in $\rho_B(E)$ are all only made of stress tensor one can directly compute the Renyi entropies of subsystem $B$ order by order in $R/\lambda_T$. 
In \cite{Dymarsky:2015cxp} we compute the vacuum subtracted Renyi entropies in our universal density matrix and find
\bea
&\Delta S_n(\rho_B(E))=\frac{(1+n)}{12n}(R/\lambda_T)^2-\frac{(1+n)}{120cn}(R/\lambda_T)^4\frac{(n^2+11)}{12n^2}\nn\\
&+\frac{(1+n)}{630c^2}(R/\lambda_T)^6\frac{(4-n^2)(n^2+47)}{144n^4}+\cdots
\eea
which is to be compared to the Renyi entropies in the thermal reduced densitry matrix $\rho_B(T)=\Tr_{B^c}(e^{-\beta H}/Z)$:
\bea
&\Delta S_n(\rho_B(T))=\frac{(1+n)}{12n}(R/\lambda_T)^2-\frac{(1+n)}{120cn}(R/\lambda_T)^4\nn\\
&+\frac{(1+n)}{630c^2}(R/\lambda_T)^6+\cdots
\eea
We observe that while for $n>1$ the Renyi entropies do not match, entanglement entropies $(n=1)$  match perfectly. 
Discrepancy between Renyi entropies is a consequence of infinite dimensionality of the Hilbert space. The Fannes-Audenaert inequality and its generalizations for $n>1$ restrict the difference between entropies to be bound by the trace distance multiplied by a factor proportional to the dimension of the Hilbert space. As the latter diverges, entanglement and Renyi entropies could be different for arbitrarily close $\rho_B(E)$ and  $\rho_B(T)$.
Equivalence of entanglement entropies though can be further used to show that
\bea
&\|\rho_B(E)-\rho_B(T)\|^2\leq S(\rho_B(E)\|\rho_B(T))\nn\\
&=\Delta \la H(\rho_B(T))\ra-\Delta S=e^{-O(S(E))}.
\eea
Here $H(\rho_B(T))=-\log \rho_B(T)$ is a local integral over $T_{00}$ in $1+1$-dimensions, and we have tuned the two states to have the same energy density. Therefore, $\Delta \la H(\rho_B(T))\ra=0$ by construction.
For a discussion of ETH in $1+1$-dimensional CFT at large central charge see \cite{Fitzpatrick:2015qma}.

%

\section{Discussions}

In this paper, we have argued that {\it chaotic} CFTs are special in that their reduced density matrix on any finite subsystem of arbitrary shape in energy eigenstates are well-approximated by a universal density matrix. In order to prove this we assumed local ETH. All integrable models we checked (free theories and minimal models in $1+1$-dimensions) failed to satisfy our local ETH assumption. Therefore, we can interpret the local ETH assumption as our working definition of chaos in CFTs.  It would be interesting to connect the local ETH to more standard definitions of quantum chaos in the literature. In particular, one might hope to use the exponential decay of out-of-time order correlators and bootstrap equations to prove a statement similar to local ETH. 


\section{Acknowledgements}
We would like to thank John Cardy, Thomas Faulkner, Liam Fitzpatrick, Daniel Harlow, Thomas Hartman, Tarun Grover, Mark Srednicki and Sasha Zhiboedov for valuable discussions. The research of NL is supported in part by funds provided by MIT-Skoltech Initiative. This paper has the preprint number Technical Report MIT-CTP/4841. This work is supported by the Office of High Energy Physics of U.S. Department of Energy under grant Contract Number  DE-SC0012567.

\bibliographystyle{apsrev4-1}


\section{Supplementary material}\label{supmat}

\section{Rindler space: convenient conformal frame}\label{AppA}

Consider a $(d+1)$-dimensional CFT in radial quantization with a ball-shaped subsystem of angular size $\theta_0$ on $S^d$ at $r=1$. According to the operator/state correspondence the density matrix in the subsystem is given by a path-integral over the $(d+1)$-dimensional space with two operators inserted, $\Psi$ at $r=\ep$ and $\Psi^\dagger$ at $r=1/\ep$ with $\ep\to 0$, and a cut open at the location of the subsystem. The initial metric in the radial quantization is 
\bea
ds^2=dr^2+r^2d\Omega_d^2
\eea
with $(\theta_1, \cdots \theta_d)$ the coordinates on $S^d$.
We perform the following conformal transformation
\bea
&&\frac{L(r^2-1)}{1+r^2+2r\cos\theta_1}=\frac{X^0}{1-2 X^1+X\cdot X}\nn\\
&&\frac{2L r \sin\theta_1\cos\theta_2}{1+r^2+2r\cos\theta_1}=\frac{(1-X\cdot X)/2}{1-2 X^1+X\cdot X}\nn\\
&&\frac{2L r \sin\theta_1\sin\theta_2\cdots \sin\theta_i}{1+r^2+2r\cos\theta_1}=\frac{X^i}{1-2 X^1+X\cdot X},\quad i>1\nn\\
&&L=\frac{1}{2}\cot(\theta_0/2).
\eea
that maps the subsystem at $r=0$ and $\theta_1\leq \theta_0$ to the negative half-space, i.e. $(0,X^1<0,0\cdots 0)$. Here $L$ is the radius of $S^d$ in units where $R$ is set to one. The new metric in the $X$-coordinates that we call Rindler frame is given by
\bea
&&ds^2=\Lambda(X)^2 dX^i dX^i\nn\\
&&\Lambda(X)=\lb X^0-\frac{L V_-}{2}-\frac{V_+}{8L}\rb^{-1}\nn\\
&&V_\pm=(1\pm 2X^1+X\cdot X).
\eea

 In these coordinates the path-integral without operator insertions prepares the Rindler density matrix in vacuum. The operators $\Psi$ and $\Psi^\dagger$ are now inserted at $X_-$ and $X_+$ respectively.
 \bea
 &&X_\pm=(\pm \sin\theta_0,\cos\theta_0,0\cdots, 0),\nn\\
 &&\Lambda(X_-)=(2\sin\theta_0)^{-1},\nn\\
 &&\Lambda(X_+)=\ep^{-2}(2\sin\theta_0)^{-1}.
 \eea
 Under this map a conformal primary transforms according to
 \bea
\la  \Psi(r=0)\cdots \ra_{\Lambda(X)\delta_{ij}}=\Lambda(X(r=0))^{-h}\la \Psi(X(r=0)\cdots \ra_{\delta_{ij}}\nn
 \eea
Therefore,
\bea
\la\Psi(1/\ep)\Psi(\ep)\cdots \ra_{radial}=(2\ep\sin\theta_0)^{2h}\la \Psi(X_+)\Psi(X_-)\cdots \ra_{Rind}\nn
\eea
 In the thermodynamic limit $\theta_0\ll 1$ the distance between $\Psi$ and $\Psi^\dagger$ goes to zero: $|X_+-X_-|=2\sin\theta_0\ll 1$, and we use the OPE to obtain
 \bea
 &&\la\Psi(1/\ep)\Psi(\ep)\cdots \ra_{radial}=\ep^{2h}\sum_p  (2\sin\theta_0)^{h_p}\la \mO_p(X_0)\cdots\ra\nn.
 \eea
 

\section{Spinless descendant Eigenstates}\label{AppB}
\subsection{Local probes}
An arbitrary descendant energy eigenstate in conformal field is created by the operator $P_{\mu_1}\cdots P_{\mu_n}\Psi_{\nu_1\cdots \nu_m}$. In  order to simplify the presentation and avoid unnecessary manipulation of indices we focus on a particular class of spinless primaries: $(P^2)^l\Psi_a$. The argument generalizes to arbitrary descendants. Our eigenstates of interest are labelled by $(a,l)$. In a conformal theory the matrix element of a scaling operator $\mO_p$ in these states is given by 
\bea\label{1ptoff-diag}
&&\la E_{(b,l)} |\mO_p|E_{(a,m)}\ra\nn\\
&&=\frac{1}{L^{h_p}}\frac{\la\Psi_b(\infty)(K_{\mu}K^\mu)^l\mO_p(1)(P_\nu P^\nu)^m\Psi_a(0)\ra}{\sqrt{\la X_{a,m} \ra\la X _{b,l}\ra}},
\eea
where $\la X_{a,m}\ra=\la\Psi_a(\infty)(K_\mu K^\mu)^m(P_\nu P^\nu)^m\Psi_a(0)\ra$.
Note that in radial quantization $(P_\mu|\Psi\ra)^\dagger=\la \Psi|K_\mu$. Here, we assume that $l$ and $m$ are much smaller than $h_a$ and $h_b$. 

First, consider the term $\la X_{a,m}\ra$ in the denominator. Primary field $\Psi_a$ is killed by $K_\mu$, so we only need to compute $[(K_\nu K^\nu)^m,(P_\mu P^\mu)^m]\Psi_a$. 
This operator can be simplified by the successive application of the following commutation relations of the generators of the conformal group: 
\bea
&&[K_\mu,P_\nu]=2(\delta_{\mu\nu}D-M_{\mu\nu})\nn\\
&&[M_{\mu\nu},P_\rho]=P_{[\mu}\delta_{\nu]\rho}\nn\\
&&[M_{\mu\nu},K_\rho]=K_{[\mu}\delta_{\nu]\rho}\nn\\
&&[M_{\mu\nu},M_{\rho\sigma}]=M_{\rho][\mu}\delta_{\nu][\sigma}.
\eea
Every time the operator $D$ appears it acts on its eigenstate and we get a factor $h_a+i$, where $0\leq i\leq m$. All other terms carry no powers of $h_a$. At large $h_a$ we find
\bea
\la X_{a,m}\ra=(d+1)^m (2 h_a)^{2m} (1+O(h_a^{-1})).
\eea
The numerator in (\ref{1ptoff-diag}) has two terms
\bea\label{2terms}
&&\la\Psi_q \mO_p (K_{\mu}K^\mu)^l(P_\nu P^\nu)^m\Psi_s(0)\ra\nn\\
&&+\la \Psi_q [(K_\mu K^\mu)^l,\mO_p](P_\nu P^\nu)^m\Psi_s(0)\ra
\eea
With no loss of generality we assume $m\geq l$. Here we argue that the first term wins over the second term in the large $h_a$ limit. As we saw above, the first term at large $h_a$ scales as
\bea
&&(d+1)^l (2h_s)^{2l}\la \Psi_q(\infty) \mO_p(1) (P_\nu P^\nu)^{m-l}\Psi_s(0)\ra\nn\\
&&=(d+1)^l (2h_s)^{2l}\lim_{z\to 0}(\p^2)^{m-l}_z\la \Psi_q(\infty)\mO_p(1)\Psi_s(z)\ra\nn\\
&&=(d+1)^l (2h_s)^{2l} C^p_{qs} f_{m-l}(h_p)
\eea
where $f_{m-l}(h_p)$ is a polynomial of degree $2(m-l)$ in $h_p$. To compute the second term in (\ref{2terms}) we work out the commutator
\bea\label{commutator}
&&[(K_\mu K^\mu)^l,\mO_p]=l [K_\mu,\mO_p] K^\mu (K_\alpha K^\alpha)^{l-1} \nn\\
&&+l (l-1) [K_\mu,[K_\nu,\mO_p]]K^\mu K^\nu(K_\alpha K^\alpha)^{l-2} \nn\\
&&+\cdots +[K_{\mu_1},[K_{\mu_2},\cdots [K_{\mu_l},\mO_p]]\cdots ]K^{\mu_1}\cdots K^{\mu_l}.\nn
\eea
The special conformal transformation generated by vector field $\ep \xi^\mu$ sends
\bea
x'^\mu=\frac{x^\mu-\ep \xi^\mu x^2}{(1-2\ep \xi_\mu x^\mu+\ep^2 \xi^2 x^2)}
\eea
which transforms the scaling operator $\mO_p$ according to
\bea
(1-2\ep \xi_\mu x^\mu+\ep^2\xi^2 x^2)^{h_p}\mO(x'^\mu)=e^{-i\ep \xi_\mu K^\mu}\mO(x)e^{i\ep \xi_\nu K^\nu}.\nn
\eea
Matching the coefficients of $\frac{\ep^n}{n!}$ in a series expansion on both sides gives
\bea
&&[K_{\mu_1},[K_{\mu_2}\cdots [K_{\mu_n},\mO_p]\cdots ]\nn\\
&&=\p_\ep^n\lb(1-2\ep \xi_\mu x^\mu+\ep^2 a^2 \xi^2)^{h_p}\mO(x'^\mu) \rb\nn
\eea
Terms that appear on the right-hand side of the equation above have the form
\bea
f(x^\mu,\xi^\mu,h_p)(\p\cdots \p\mO_p)K_{\mu_1}\cdots K_{\mu_j}.
\eea
Putting this back in the second term in (\ref{2terms}) we obtain terms that are
\bea
g(h_p)(d+1)^j (2h_a)^{2j}\la \Psi_b (\p\cdots \p\mO_p)(1) \Psi_a\ra
\eea
for $j<l$. Note that extra derivatives on $\mO_p$ do not lead to any extra powers of $h_a$. As a result, the first term in (\ref{2terms}) dominates. Putting all these terms back in equation (\ref{1ptoff-diag}) we obtain the matrix elements of $\mO_p$ in energy eigenbasis
\bea
&&\la E_{(b,l)} |\mO_p|E_{(a,m)}\ra\nn\\
&&=\frac{C^p_{ab}}{L^{h_p}}(d+1)^{(l-m)/2}(2h_s)^{l-m}f_{m-l}(h_p)(1+O(h_a^{-1})).\nn
\eea
In the case $l=m$ the above expression becomes
\bea
\la E_{(b,l)} |\mO_p|E_{(a,m)}\ra=\frac{C^p_{ab}}{L^{h_p}}(1+O(h_a^{-1})).
\eea

The conformal algebra fixes their value in terms of $C^p_{ab}$, $h_a$, $h_b$ and $h_p$. In appendices, we work out these matrix elements and argue that at large $h_a$ and $h_b$ they are given by
\bea
\la E_{(b,l)} |\mO_p|E_{(a,m)}\ra=\frac{C^p_{ab}}{L^{h_p}}g_{(m-l)}(h_p) h_a^{l-m}{L^{h_p}}\lb 1+O(h_p/h_a)\rb\nn
\eea
where without loss of generality, we have assumed $m\geq l$, and $g_k(h_p)$ is a polynomial of order $2k$ in $h_p$ with $g_0(h_p)=1$. Then, from equation (\ref{eo}) we find that 
\bea
&&\la E_{(a,m)}|\mO_p|E_{(a,m)}\ra\nn\\
&&=\la E_{(b,a-b+m)}|\mO_p|E_{(b,b-a+m)}\ra \lb 1+O(h_a^{-1}) \rb\nn\\
&&=C^p_{aa}\lb 1+O(h_a^{-1}) \rb,
\eea
which together with the assumption of local ETH for primary energy eigenstates implies
\bea\label{1ptDesc}
&\la E_{(b,l)}|\mO_p|E_{(a,m)}\ra=\lb \mO_p(E)\delta_{ab}+\mO^p_{ab}\rb\lb \delta_{lm}+O(h_a^{-1})\rb.\nn \\
&
\eea

\subsection{Density matrix}
Now consider the spinless energy eigenstate $|E_{a,l}\ra$ created with a path-integral over the unit ball with $(P^\mu P_\mu)^l\Psi_a(0)=\Box^l \Psi_a(0)$ in the center of radial quantization. In the Rindler frame the Laplacian is 
\bea
\Box=\frac{1}{\Lambda(X)^{d+1}}\p_i\lb \Lambda(X)^{d-1}\p_i\rb
\eea
and the unit ball is mapped to the lower half-plane $X^0<0$. In these coordinates, the operators that create the state and its conjugate are
\bea
&&\tilde{\Psi}(X_-)=\lb \Lambda(X_\ep)^{-d-1}\p_i\lb \Lambda(X_\ep)^{d-1}\p_i\rb\rb^l \Lambda(X_\ep)^{-h}\Psi(X_\ep)\nn\\
&&\tilde{\Psi}^\dagger(X_+)=\lim_{\ep\to 0}\ep^{-2(h+2l)}\lb \Lambda(X_{1/\ep})^{-d-1}\p_i\lb \Lambda(X_{1/\ep})^{d-1}\p_i\rb\rb^l \nn\\
&&\times\:\Lambda(X_{1/\ep})^{-h}\Psi(X_{1/\ep})\nn
\eea
where we have used the fact that under conjugation $X^0\to -X^0$. Note that
\bea
&&\Lambda(X_{1/\ep})\sim \ep^{-2}, \nn\\
&&\p_i^n\Lambda(X_{1/\ep})\sim \ep^{-(2+n)}
\eea
and since $\p_i\Psi$ will not carry any powers of $\ep$ the conjugate operator has the form
\bea
\tilde{\Psi}(X_+)=f(L)\Psi(X_+).
\eea
Therefore, the density matrix in these coordinates is
\bea
tr(\rho\cdots )=\frac{\la\tilde{\Psi}(X_-)\Psi(X_+)\cdots\ra}{\la \la\tilde{\Psi}(X_-)\Psi(X_+)\ra}
\eea
The operator $\tilde{\Psi}\Psi$ simplifies further in the thermodynamic limit $L\gg 1$. To see this, apply $\p_i$ to the OPE
\bea
&&\frac{\p}{\p X^i}\Psi(X)\Psi(X_+)=\sum_p \frac{\p}{\p X^i}\lb (X-X_+)^{h_p-2h}\mO(X_+)\rb\nn\\
&&=(-2h) (X-X_+)^{h_p-2h-1} \mO(X_+)+O(1/h),
\eea
where we have used $h\gg h_p$. Now, notice that
\bea
&&\Lambda(X_\ep)\sim L,\nn\\
&& \p_i^n\Lambda(X_\ep)\sim L^{n+1}.\nn
\eea
However when $\p_i$ acts on $\Lambda(X_\ep)$ we get a factor of $h L$. Therefore, 
\bea
\tilde{\Psi}(X_-)\Psi(X_+)=\Lambda(X_-)^{-(h+2l)}\lb (\p_i\p_i)^{2l}\Psi\rb(X_-)\Psi(X_+).\nn
\eea
Finally, the expression for the density matrix in these coordinates becomes
\bea
&&\frac{\sum_p(2/L)^{h_p}C^p_{\psi\psi}(\p_i\p_i)^l(z-X_+)^{-2h}\big|_{z\to X_-}\la \mO_p\cdots \ra}{(\p_i\p_i)^l(z-X_+)^{-2h}\big|_{z\to X_-}}\nn\\
&&=\sum_p(2/L)^{h_p}C^p_{\psi\psi}\la \mO_p\cdots\ra
\eea
which is the same as the density matrix in the primary state from which $|E_a\ra$ descends. As a result, we find 
\bea\label{descendant}
\|\rho_B^{(a,m)}-\rho_B^a\|\sim O\lb(E_aL)^{-1}\rb.
\eea

\section{Subsystem ETH implies local ETH}\label{AppE}
Consider an observable $A=\sum_a a|a\ra\la a|$ and the operator $\rho-\rho_T$ in this basis:
\bea
\rho-\rho_T=\sum_{ab}c_{ab}|a\ra\la b|.
\eea
The expectation value of $A$ is
\bea\label{upper}
&&\Tr ((\rho-\rho_T)A)=\sum_a c_{aa}\: a\leq \sum_a |c_{aa}|\:a\nn\\
&&\leq \lb\sum_a |c_{aa}|\rb^{1/2} \lb \sum_a|c_{aa}| a^2\rb^{1/2}\nn\\
&&\leq \|\Phi[\rho-\rho_T]\|^{1/2}\lb \sum_a|c_{aa}| a^2\rb^{1/2}\nn\\
&&\leq \|\rho-\rho_T\|^{1/2}\lb \sum_a|c_{aa}| a^2\rb^{1/2},
\eea
where $\Phi[\rho]=\sum_a\rho_{aa}|a\ra\la a|$ is the map that decoheres $\rho$ in the basis of $A$, and we have used the fact that this map decreases the trace distance of operators. Note that $c_{aa}$ are real, but could have either sign. Denote by $V$ and $W$ the projectors that project to $c_{aa}$ that are, respectively, positive and negative.
 Then,
\bea
&&\sum_a|c_{aa}| a^2=\sum_{a\in V}\Tr\lb V(\rho-\rho_T)V A^2\rb\nn\\
&&-\sum_{a\in W}\Tr\lb W(\rho-\rho_T)W A^2\rb\nn\\
&&\leq \Tr\lb (\rho+\rho_T)A^2\rb,
\eea
where we used the fact that for a projector $V$: $\Tr (V \rho V A^2)\leq \Tr (\rho A^2)$.
Putting this back into (\ref{upper}) we find
\bea
\Tr ((\rho-\rho_T)A)\leq \|\rho-\rho_T\|^{1/2}\Tr\lb (\rho+\rho_T)A^2\rb^{1/2}
\eea
Repeating the argument above for $\tilde{\rho}_B^{a,b,\alpha}-\frac{1}{2}(\tilde{\rho}_B(E_a)+\tilde{\rho}_B(E_a))$ we find
\bea
&\Tr ((e^{i\alpha}\sigma_{ab}+e^{-i\alpha}\sigma_{ba})A)\leq  \|e^{i\alpha}\sigma_{ab}+e^{-i\alpha}\sigma_{ba}\|^{1/2}\times\nn\\
& \Tr \lb (\tilde{\rho}_B^{a,b,\alpha}+\frac{1}{2}(\tilde{\rho}_B(E_a)+\tilde{\rho}_B(E_a))A^2\rb^{1/2}.\nn
\eea
for all $\alpha$. By adding and subtracting the inequality above for values of $\alpha=0$ and $\alpha=\pi$ we find $\Tr (A \sigma_{ab})=e^{-O(S(E))}$.

\begin{figure}[t]
\centering
\includegraphics[width=0.4\textwidth]{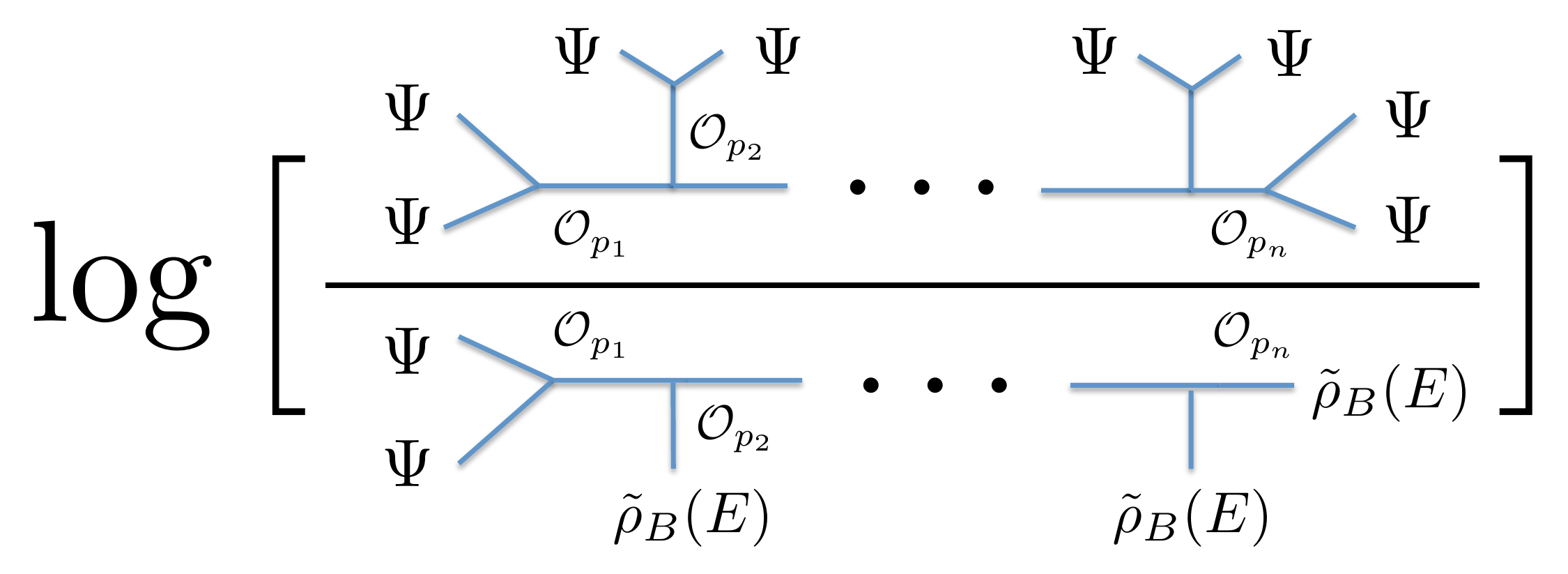}\\
\caption{\small{The relative entropy replica trick.}}
\label{fig4}
\end{figure}

\section{Relative entropy replica trick}
Similar to entanglement entropy there is a replica trick that computes the relative entropy of arbitrary states in quantum field theory \cite{Lashkari:2014yva,Lashkari:2015dia}. The relative entropy is found from the analytic continuation in $n$ of
\bea
S(\tilde{\rho}^a_B\|\tilde{\rho}_B)=\lim_{n\to 1}\frac{1}{n-1}\log\left[ \frac{{\rm Tr}((\tilde{\rho}_B^a)^n){\rm Tr} (\tilde{\rho}_B)^{n-1}}{{\rm Tr}(\tilde{\rho}_B^a\tilde{\rho}_B^{n-1}){\rm Tr}(\tilde{\rho}_B^a)^{n-1}}\right]\nn
\eea
Now inserting~\eqref{11} into the above expression we find that 
\bea\label{entropyexpan}
&&S(\tilde{\rho}^a_B\|\tilde{\rho}_B)=\p_n\log \left[\frac{1+n A^{(1)}_n+n A^{(2)}_n+\cdots }{1+A^{(1)}_n}\right]_{n\to 1}\nn\\
&&A^{(1)}_n=\frac{{\rm Tr}(\cR_B^a \tilde{\rho}_B^{n-1})}{{\rm Tr}(\tilde{\rho}_B^n)}\nn\\
&&A^{(2)}_n=\sum_{i=1}^{n-2}\frac{\Tr \lb \cR_B^a \tilde{\rho}_B^i\cR_B^a \tilde{\rho}_B^{n-2-i} \rb}{{\rm Tr}(\tilde{\rho}_B^n)}\nn\\
&&A^{(k)}_n=\sum_{m_1+\cdots m_k=n-k}\frac{\Tr \lb \cR_B^a \tilde{\rho}_B^{m_1}\cR_B^a \tilde{\rho}_B^{m_2}\cdots  \cR_B^a \tilde{\rho}_B^{m_k} \rb}{{\rm Tr}(\tilde{\rho}_B^n)}\nn
\eea
Consider the term $A^{(k)}$. From the definition (\ref{rhouniv}) we find
\bea
&&A^{(k)}_n=\sum_{m_1+\cdots m_k=n-k}\sum_{p_1,\cdots p_k}\lb O^{p_1}_{aa}\cdots O^{p_k}_{aa}\rb f^{p_1\cdots p_k}_{m_1\cdots m_k}R^{h_{p_1}+\cdots h_{p_k}}\nn\\
&&f^{p_1\cdots p_k}_{m_1\cdots m_k}=\frac{\Tr \lb \mO_{p_1} \tilde{\rho}_B^{m_1} \mO_{p_2} \tilde{\rho}_B^{m_2}\cdots   \mO_{p_k} \tilde{\rho}_B^{m_k} \rb}{{\rm Tr}(\tilde{\rho}_B^n)}
\eea
It is clear that $\lb O^{p_1}_{aa}\cdots O^{p_k}_{aa}\rb=O(\eta_a^k)$ where $\eta_a=\sup_pO^p_{aa}=e^{-O(S(E))}$. Therefore, if we argue that $ f^{p_1\cdots p_k}_{m_1\cdots m_k}$ is not entropically suppressed we have argued $A^{(k)}=O(\eta_a^k)$. Using (\ref{rhouniv}) we have
\bea
&&f^{p_1\cdots p_k}_{m_1\cdots m_k}=\nn\\
&&\frac{\sum_{q_1\cdots q_k}\lb O_{q_1}(E)\cdots O_{q_k}(E)\rb  \la \mO_{p_1}\mO_{q_1}^{m_1}\cdots \mO_{p_k}\mO_{q_k}^{m_k}\ra R^{h_{p_1}+\cdots h_{p_k}}}{\sum_{q_1\cdots q_n}\lb O_{q_1}(E)\cdots O_{q_n}(E)\rb \la \mO_{q_1}\cdots \mO_{q_n}\ra R^{h_{q_1}+\cdots h_{q_n}}}\nn
\eea
 Neither the correlators, nor $O_p(E)$ have any entropic suppressions. Therefore, as long as the correlators on the $n$-sheeted manifold are finite, the sums are convergent and $A^{(k)}=O(\eta_a^k)$. Therefore, expanding (\ref{entropyexpan}) in $\eta_a$ we have
 \bea
 S(\tilde{\rho}^a_B\|\tilde{\rho}_B)&=&\p_n\lb (n-1)A^{(1)}_n\rb_{n\to 1}+O(\eta_a^2)\nn\\
 &=&O(\eta_a^2)=e^{-O(S(E))}.
 \eea

\end{document}